\documentclass[aps,prb,twocolumn,floatfix,showpacs]{revtex4-1}
\usepackage{graphicx}
\usepackage{amsmath}
\usepackage{bm}
\usepackage{verbatim}

\usepackage[colorlinks=true,citecolor=blue]{hyperref}
\hypersetup{colorlinks=true,citecolor=blue,linkcolor=red,urlcolor=blue}

\begin{document}
\title{Boltzmann conductivity of ferromagnetic graphene with magnetic impurities}

\author{B. Zare Rameshti}
\affiliation{Department of Physics, Institute for Advanced Studies
in Basic Sciences (IASBS), Zanjan 45137-66731, Iran}
\author{A. G. Moghaddam}
\email{agorbanz@iasbs.ac.ir}
\affiliation{Department of Physics,
Institute for Advanced Studies in Basic Sciences (IASBS), Zanjan
45137-66731, Iran}
\author{S. H. Abedinpour}
\affiliation{Department of Physics, Institute for Advanced Studies
in Basic Sciences (IASBS), Zanjan 45137-66731, Iran}
\author{S. Abdizadeh}
\affiliation{Department of Physics, Institute for Advanced Studies
in Basic Sciences (IASBS), Zanjan 45137-66731, Iran}
\author{M. Zareyan}
\affiliation{Department of Physics, Institute for Advanced Studies
in Basic Sciences (IASBS), Zanjan 45137-66731, Iran}

\begin{abstract}
We investigate the electrical conductivity of spin-polarized graphene in the presence of short-ranged magnetic scatterers within the relaxation time approximation and the semi-classical Boltzmann approach. Spin-flip scattering of the itinerant electrons from the majority spin sub-band into the minority one results in a minimum in the electrical resistivity at a finite temperature.
While this behavior is reminiscent of the renowned Kondo effect, it has an entirely different origin and differs from the Kondo effect in several aspects. In particular, unlike the Kondo effect, this is a single particle phenomena, and it does not require antiferromagnetic coupling between the magnetic moments of impurities and spins of the itinerant electrons.
\end{abstract}

 \maketitle
\section{Introduction}\label{sec:intro}
A huge activity in fundamental and applied physics and
even chemistry of graphene has been triggered in the last few years
due to its peculiar electronic, mechanical, optical
and chemical properties~\cite{neto_rmp_2009,Novoselov-Nat-2005,Zhangt-Nat-2005,Geim-Sc-2009,Katsnelson,Katsnelson-2006}. In particular, the
excitations in graphene behave in a similar way as massless Dirac
fermions which leads to many intriguing phenomena in its
electronic properties. Some of these effects which have been
experimentally confirmed are new types of the quantum Hall effect, Klein tunneling, and non-vanishing metallic conductivity at
neutrality point.
Electronic transport in graphene in particular has been a central subject of interest~\cite{sarma_rmp_2011,SAdam,SAdam2,Nomura,Min,Novikov,Tan,Aleiner,Hwang}.
In early studies of graphene's conductivity, neutral short-ranged scatterers were considered and the conductivity had been predicted to be independent of the carrier density. On the other hand linear dependance of the conductivity on the carrier density was observed in experiments. Inclusion of long-ranged charged scatterers was essential for an adequate description of these experimental observations (for a throughout review of the subject, see, Ref.~[\onlinecite{sarma_rmp_2011}] and references therein).
The effect of magnetic scatterers on charge transport has also been investigated, mainly in the context of the Kondo effect~\cite{fritz_rpp_2013,Uchoa11,chen_nphys_2011}. Magnetic scattering centers usually originate from a magnetic ad-atom or a lattice vacancy, but also from an edge effect. Graphene with dilute magnetic doppings have been also explored for magnetic ordering and for the so called RKKY coupling of impurities~\cite{Vozmediano05,Abanin10,Black-Schaffer10,Sherafati11,Kogan11,Power13}.

Recently the potential of graphene for spintronics applications
has been revealed since it shows long spin relaxation lengths of
few microns at room temperature\cite{grapheneFerroMetal1}. Subsequently, the interest in
studying magnetic properties of graphene in one hand and transport
properties of magnetized graphene on the other hand has increased.
Magnetized graphene or more precisely graphene with spin imbalance
has been suggested to be realized in a variety of ways. Some
theoretical studies predict intrinsic ferromagnetic correlations
can exist in graphene sheets~\cite{Peres05} and nanoribbons~\cite{RibonFerro}. Another
possible way is to use an insulating ferromagnetic substrate or
alternatively add a magnetic material or magnetic impurities on
top of the graphene sheet\cite{grapheneFerroMetal2,grapheneFerroMetal3,InsulFerro}.
Thanks to the gapless excitation spectrum of graphene, and fine tunability of its chemical potential through external gates, the splitting energy between up and down spin carriers in spin-polarized graphene could be made comparable, or even larger than the chemical potential. In this regime majority and minority spin electrons can manifestly become of electron and hole like nature, respectively. Several properties of spin-polarized graphene in this \emph{spin-chiral} regime have been already explored by some of us~\cite{malek_prb_2008, ali_prb_2008, ali_prl_2010,Vahid,fariborz_prb_2013,Babak}.

In this work we study the electrical conductivity of spin-polarized graphene considering short-ranged magnetic impurities as the sole source of scattering and resistance. We use semiclassical Boltzmann formalism and the relaxation time approximation. As the density of states (DOS) of graphene linearly depends on energy, minority and majority spin electrons in spin-polarized graphene will clearly have different DOS. In particular, when the spin splitting is equal to the chemical potential, the DOS of minority spin carriers vanishes. This will have remarkable effects on the rates of spin-flip scatterings from majority to minority (and vice versa) spin sub-bands. As a result, a minimum at finite temperature in the total resistivity will appear.  While this feature resembles the well known Kondo effect, the underlying physics of these two phenomena are totally different.
\par
This article is organized as follows. In Sec.~\ref{sec:model}, we introduce our model and explain the method we use to calculate the resistivity in the presence of magnetic impurities within the Boltzmann method and the relaxation time approximation. In Sec.~\ref{sec:results}, we present and discuss our numerical results for the resistivity of spin polarized graphene. Finally, in Sec.~\ref{sec:concl}, we conclude and summarize our main findings.

\section{Model and Formalism}\label{sec:model}

We consider a spin-polarized graphene sheet which can be described
by the following Hamiltonian at low energies near Dirac points
(${\bf K}$ or ${\bf K^{\prime}}$),
\begin{equation} \label{b1}
{\cal H}_{0}=v_{\rm F}\hat{s}_{0}\otimes\hat{\bm \sigma}\cdot{\bf p}-h\hat{s}_{z}\otimes\hat{\sigma}_{0} ~,
\end{equation}
with Fermi velocity $v_{\rm F}$, momentum ${\bf p}=(p_x,p_y)$,
exchange splitting $h>0$, and $\bm{ \hat{\sigma}}
=(\hat{\sigma}_{x}, \hat{\sigma}_{y})$. The Pauli matrices
$\hat{\sigma}_i$ and $\hat{s}_i$ with $i=0,..,3$ operate on
pseudo-spin space (characterized by two different trigonal
sub-lattices $A$ and $B$ of the hexagonal structure of graphene),
and spin, respectively. Moreover, $\hat{\sigma}_{0}$ and $\hat{s}_{0}$
represent unit matrices of corresponding spaces,
and $\otimes$ refers to the direct Kronecker product of two spaces. Diagonalizing
the Hamiltonian~(\ref{b1}) we end up with the following
eigenstates
\begin{eqnarray}
\begin{split}
&\psi^{\dagger}_{{\bf
k}\alpha\uparrow}=\dfrac{1}{\sqrt{2}}\begin{pmatrix} \alpha
e^{i\phi_{\bf k}} & 1 & 0 & 0
\end{pmatrix}~,\\
&\psi^{\dagger}_{{\bf
k}\alpha \downarrow}=\dfrac{1}{\sqrt{2}}\begin{pmatrix} 0 & 0 &
\alpha e^{i\phi_{\bf k}} & 1
\end{pmatrix}~,
\end{split}
\label{b2-0}
\end{eqnarray}
corresponding to the eigenvalues
\begin{equation}
\varepsilon_{{\bf
k}\alpha s}=\alpha\hbar v_{\rm F}\vert{\bf k}\vert -s h~~~~~(s=\uparrow,\downarrow)~,
\label{b2}
\end{equation}
where $\alpha =\pm$ refers to conduction (+) and
valence (-) band excitations, and $\phi_{\bf
k}=\arctan\left(k_{y}/k_{x}\right)$ indicates the propagation angle.

The effect of magnetic impurities can be taken into account via a
short range interaction
between the spin of itinerant electrons $\hat{\bf
s}_{e}$ and the local moment of impurity $\hat{\bf S}({\bf r}=0)$ given by~\cite{Zener},
\begin{equation}
{\cal H}_{\rm s-d}=\frac{J}{N}\hat{\bf S}\cdot \hat{\bf
s}_{e}\otimes\hat{\sigma}_{0}~.
\end{equation}
Here $J$ is the exchange integral between itinerant electrons and
the electrons of the localized impurities and $N$ is the total
number of atoms in the crystal. If the concentration of magnetic
impurities is very low, we can use the single impurity approximation
where the interaction between the impurities is neglected. Then, the interaction between spin of the conduction electrons and spin
of the magnetic impurities can be modeled by the single impurity $s-d$ Hamiltonian,
\begin{eqnarray}\label{b13}
&&V=\dfrac{J}{N}\sum_{{\bf k}{\bf k}^{\prime},
\alpha\alpha'}\left[ S^{+} c^{\dagger}_{{\bf k}^{\prime}\alpha'
\downarrow}c_{{\bf k}\alpha\uparrow}
+S^{-} c^{\dagger}_{{\bf k}^{\prime}\alpha' \uparrow}c_{{\bf k}\alpha\downarrow}\right.\nonumber\\
&&~~~~~~\quad\left.\qquad+S^{z}( c^{\dagger}_{{\bf
k}^{\prime}\alpha' \uparrow}c_{{\bf k}\alpha\uparrow}-
c^{\dagger}_{{\bf k}^{\prime}\alpha' \downarrow}c_{{\bf
k}\alpha\downarrow})\right] ~,
\end{eqnarray}
where, $c_{{\bf k}\alpha s}$  ($c^{\dagger}_{{\bf
k}\alpha s}$) destroys (creates) an electron
with momentum $k$ and spin $s$ in the $\alpha$-band of
 graphene, and $S^\pm= S^x\pm i S^y $.
The first two terms in Eq. (\ref{b13})
are responsible for the spin-flip scatterings while the last term
is responsible for the spin-conserving scatterings. We assume
that all the magnetic impurities have the same moment $M$.
\par
In order to calculate the resistivity in the presence of magnetic
impurities, we use the Boltzmann transport theory in the
relaxation time approximation scheme. The relaxation time for each
process which can be either spin-conserving or spin-flipping, can
be found from the relation
\begin{eqnarray}\label{tau}
\dfrac{1}{\tau_{ s\alpha,s'\alpha'}({\bf k})}=\int \dfrac{d{\bf
{\bf k}}^{\prime}}{(2\pi)^2} W_{{\bf k}\alpha s, {\bf
k}^{\prime}\alpha' s'} \left(1-\cos\phi_{{\bf k}{\bf
k}^{\prime}}\right)~,
\end{eqnarray}
where $\phi_{{\bf k}{\bf k}^{\prime}}=\phi_{{\bf
k}^{\prime}}-\phi_{{\bf k}}$ is the angle between
scattering and incidence directions, and
the scattering rates
$W_{{\bf k}\alpha s, {\bf k}^{\prime}\alpha' s'}$ in the
first-order Born approximation are given by the Fermi's Golden rule
\begin{equation}
W_{{\bf k}\alpha s, {\bf k}^{\prime}\alpha' s'
}=\dfrac{2\pi}{\hbar}n_{\rm imp} \left\vert T_{{\bf k}\alpha s,
{\bf k}^{\prime}\alpha' s'}\right\vert^{2}
\delta\left(\varepsilon_{{\bf k}\alpha s}-\varepsilon_{{\bf
k}'\alpha' s'}\right)~,
\end{equation}
with $n_{\rm imp}$ being the density of magnetic impurities. The
scattering probabilities follow from the $s-d$  potential of magnetic impurities,
\begin{equation}
T_{{\bf k}\alpha s, {\bf k}'\alpha's'} =
\langle {\bf k}'\alpha's'\vert\otimes\langle M'_{ z}\vert V \vert M_{z}\rangle\otimes\vert {\bf k}\alpha s\rangle~,
\end{equation}
in which we choose the basis as the multiplication of conduction
electron state $\vert {\bf k} \alpha s\rangle$ and impurity spin
state $\vert M_{z}\rangle$. We drop the indices $M_z$ and $M'_z$
in $T_{{\bf k}\alpha s, {\bf k}'\alpha's'} $ since for any $M_z$
the final state of impurity spin $M'_z$ is determined with the
incident and scattered electrons spin $s$ and $s'$. In fact, for
spin-conserving processes $M'_z=M_z$, and for spin-flip process
$M'_z=M_z\pm1$ for $s=-s'=\uparrow$ and $s=-s'=\downarrow$,
respectively.
\par
Note that, in a spin-conserving scattering \emph{i.e.}, $s'=s$, energy conservation implies $\alpha'=\alpha$, while in a spin-flip scattering \emph{i.e.}, $s'=-s$, depending on the energy of scattered electron $\varepsilon$,  either $\alpha'=\alpha$ (for $|\varepsilon|>h$) or $\alpha'=-\alpha$ (for $-h<\varepsilon<h$) is permitted.
\par
The amplitudes of elastic spin-conserving processes are given by $
T_{{\bf k}\alpha \uparrow, {\bf k}'\alpha \uparrow }=-T_{{\bf
k}\alpha \downarrow, {\bf k}^{\prime}\alpha \downarrow }=(J\Omega
M_z) F_{{\bf k}\alpha,{\bf k}^{\prime}\alpha}$ with $\Omega$
indicating the area of the unit cell. Similarly, the amplitudes of
spin-flip processes are $T_{{\bf k}\alpha \uparrow, {\bf
k}'\alpha' \downarrow } =A_{+} F_{{\bf k}\alpha,{\bf k}'\alpha'}$
and $T_{{\bf k}\alpha \downarrow, {\bf k}'\alpha' \uparrow }=A_{-}
F_{{\bf k}\alpha,{\bf k}'\alpha'}$, in which $A_{\pm}=J\Omega
\left[S(S+1)-M_{z}(M_{z}\pm 1)\right]^{1/2}$ and the form factor
is given by
\begin{equation}
F_{{\bf k}\alpha,{\bf
k}'\alpha'}=\dfrac{1}{2}\left[1+\alpha\alpha^{\prime}
e^{i(\phi_{{\bf k}^{\prime}}-\phi_{{\bf k}})}\right]~.
\end{equation}
Assuming randomly oriented spins for magnetic impurities we can
use the average square value $\langle M_{z}^{2}\rangle=S(S+1)/3$
instead of $M_z^2$ and then the relaxation times are obtained as,
\begin{eqnarray}
&&\dfrac{1}{\tau_{s\alpha, s\alpha }}=\pi\dfrac{\langle
M_{z}^{2}\rangle}{8\hbar} J^2\Omega^2 n_{\rm imp} \nu_{s}(\varepsilon_{{\bf k}\alpha s})\label{b4}~,\\
&&\dfrac{1}{\tau_{s\alpha, -s\alpha'}}=\dfrac{\langle
M_{z}^{2}\rangle}{2\hbar} J^2\Omega^2 n_{\rm
imp}\nu_{-s}(\varepsilon_{{\bf k}\alpha s
})B_{\alpha\alpha'}(\phi_{\bf k})~,\label{b5}
\end{eqnarray}
where $\nu_{s}(\varepsilon)=|\varepsilon+s h|/\pi(\hbar
v_{F})^{2}$ is the DOS of s-spin electrons, and $B_{\alpha\alpha'}(\phi_{\bf k})=(2\cos\phi_{\bf
k}-\pi/2)(\alpha\alpha^{\prime}-1)+\pi/2$ introduces a dependence
on the incidence angle $\phi_{\bf k}$.

We can also define the relaxation time of spin-$s$ electrons as
$\tau^{-1}_{ s \alpha}({\bf k})=\tau^{-1}_{s\alpha,s\alpha}({\bf k})+\tau^{-1}_{s\alpha,-s\alpha'}({\bf k})$,
which is based on the fact that the total rate of scattering for spin-$s$
electrons is the sum of the corresponding spin-conserving and
spin-flipping scattering rates. Then the conductivity for each
spin channel follows from,
\begin{equation}
\rho_{s}^{-1}=\sigma_{s}=-\dfrac{  e^{2}v^{2}_{F}}{2}
\sum_{\alpha}\int
\dfrac{d{\bf k}}{(2\pi)^{2}} \tau_{s\alpha}({\bf
k})\dfrac{\partial f}{\partial \varepsilon_{{\bf k}\alpha
s}}~.\label{b8}
\end{equation}
Here
$f(\varepsilon)=1/[1+e^{(\varepsilon-\mu)/(k_{\rm B} T)}]$ is the
Fermi-Dirac distribution function in which $\mu$ and $T$ indicate
the chemical potential measured from the non-magnetic state neutrality point and
temperature, respectively, and $k_{\rm B}$ is the Boltzmann constant.
Inserting the relations for relaxation times from Eqs.~(\ref{b4}) and~(\ref{b5})
into Eq. (\ref{b8}), and changing the integration variables from ${\bf
k}$ to energy $\varepsilon$ and incidence angle $\phi_{\bf k}$, we
end up with the following form for the conductivity of spin-$s$
channel,
\begin{equation}\label{sigma}
\sigma_{s}\left(k_{\rm B}T/h,\mu/h\right)=\sigma_0 \int
d\varepsilon (- \dfrac{\partial
f}{\partial\varepsilon}) \Gamma_{s}(\varepsilon)~,
\end{equation}
with
\begin{equation}\label{gamma}
\Gamma_{s}(\varepsilon)=\int d\phi \frac{
1}{1+(4/\pi)B(\phi)\nu_{-s}(\varepsilon)/\nu_{s}(\varepsilon)}~.
\end{equation}
Here $1/\sigma_{0}=\rho_0=(\pi^2 n_{\rm imp}/3\hbar
e^{2}v_{F}^{2})S(S+1)(J\Omega)^{2}$, and
$B(\phi)=(2\cos\phi-\pi/2)({\rm sgn}(|\varepsilon|-h)-1)+\pi/2$.
The relation for $B(\phi)$ follows directly from
$B_{\alpha\alpha'}(\phi_{\bf k})$ introduced above and the fact
that the band indices $\alpha$ and $\alpha'$ for two spin species
are the same for energies $|\varepsilon|>h$ and opposite to each
other when $-h<\varepsilon<h$.
\begin{figure}
\includegraphics[width=8cm]{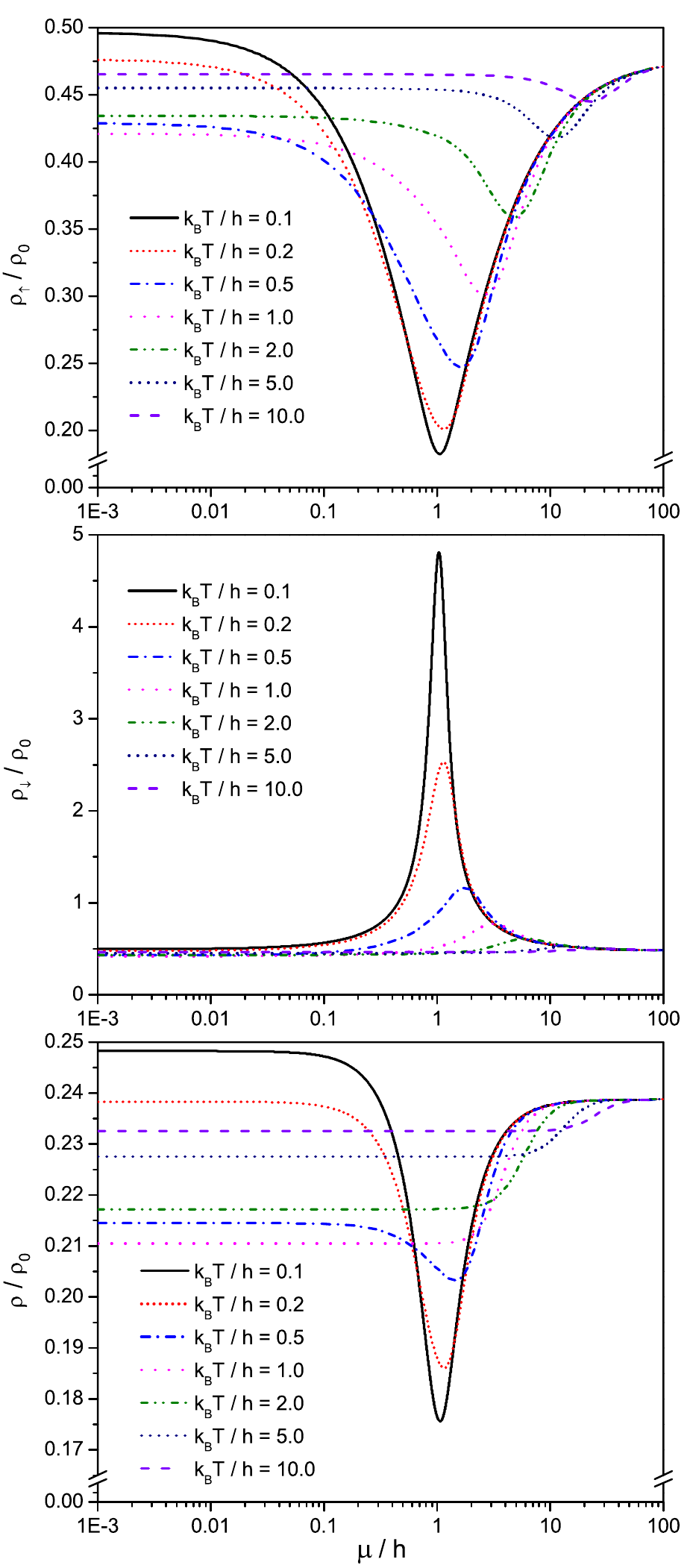}
\caption{(Color online) Current resistivity in up-spin (top panel) and down-spin (middle panel) channels, as well as the total resistivity (bottom panel) as a function of dimensionless chemical potential
$\mu/h$, for different temperatures.}\label{fig1}
\end{figure}
Now the total conductivity can be obtained from the sum of two spin
conductivities $\sigma=\sigma_{\uparrow}+\sigma_{\downarrow}$,
since the two spin channels effectively conduct the electrons
similar to the two resistors in parallel.

\section{Results}\label{sec:results}
\begin{figure}
\includegraphics[width=8cm]{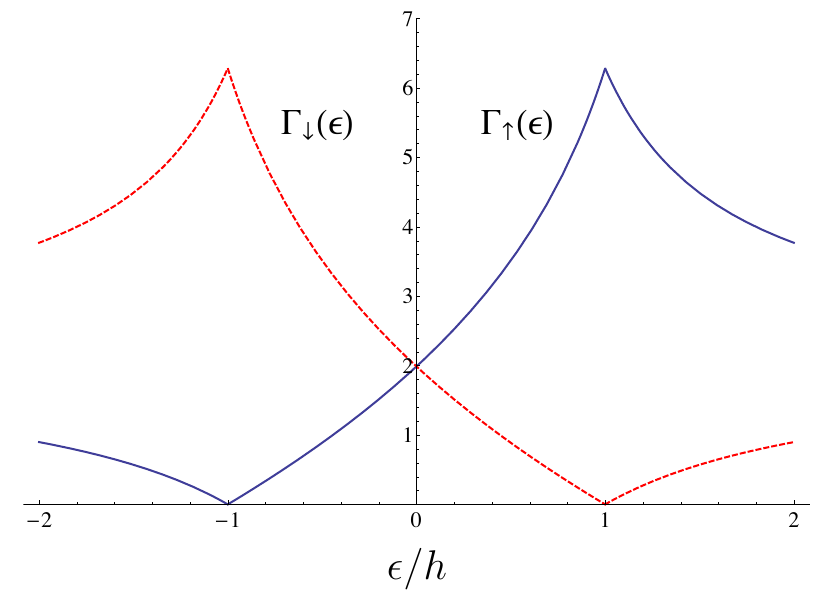}
\caption{(Color online)
The dependance of current rates $\Gamma_s$ on energy.
At $\varepsilon=s h$ the rate $\Gamma_{s}$ passes through a maxima and vanishes at $\varepsilon=- s h$.
}\label{fig2}
\end{figure}

As it is clear from Eq. (\ref{sigma}) the resistivity of the system
is a function of temperature $T$ and chemical potential $\mu$
scaled with the exchange field $h$. Figure \ref{fig1} shows the resistivity
of each spin channel and the total resistivity as functions of
$\mu/h$ for different temperatures. We first discuss the very low
temperatures when the resistivities $\rho_{s}$ directly follow
 $\Gamma_{s}(\mu)$.
The up spin resistivity $\rho_{\uparrow}$ shows a minimum at $\mu \approx h$ since at this point
the DOS for down spins vanishes and subsequently no
spin-flip scattering can happen. This leads to an increase in the
relaxation time $\tau_{\uparrow}$ and subsequent decline in the
resistivity. On the other hand resistivity of down spins
$\rho_{\downarrow}$ increases in the vicinity of $\mu=h$. This can
be understood from the fact that the density of spin down carriers
itself vanishes close to this point and as a result resistivity of down spin channel
diverges at $\mu=h$, when the temperature is close to zero.
The total resistivity, however, behaves somehow similar to the up spins' contribution and reveals a minimum close to $\mu/h=1$, where the DOS of down spins vanishes. This means that the effect of suppression of the spin-flip scattering and subsequent increase in conductivity
of up-spin channel dominates over the decline in the conductivity of spin-down channel. In order to give a more intuitive picture of this effect, we illustrate in Fig. \ref{fig2}  the energy dependence of $\Gamma_{\uparrow(\downarrow)}$ which shows a maximum at $\varepsilon=h$ ($\varepsilon=-h$) and vanishes at $\varepsilon=-h$ ($\varepsilon=h$).
As a direct result, the conductivity rises up for $\mu\rightarrow h$ and the total resistivity decreases. Far enough from the Dirac point for down spins ($\mu=h$) the resistivity reaches a constant value independent of the chemical potential.
The resistivity generally depends on temperature but becomes independent of it for $\mu\gg h$, since in this limit graphene is
practically unpolarized and $\nu_\uparrow (\varepsilon)\approx \nu_\downarrow(\varepsilon)$, then spin-flip and spin-conserving scatterings from the magnetic impurities occur with almost equal weights which are independent of energy
\begin{equation}
\Gamma_{s}(\varepsilon)=\int  \frac{d\phi}{1+(4/\pi)B(\phi)}=\dfrac{2\pi}{3}~~~(\varepsilon\gg h)~.
\end{equation}

Upon increasing the temperature not only the states
close to the Fermi level but also those activated by thermal
energy contribute to the conductivity. In general the effect of temperature
is to suppress the variation of resistivity with $\mu/h$ and the amplitude of overall change in
resistivity decreases at higher temperatures. In particular when the temperature is comparable with exchange splitting $h$ the
resistivity changes drastically. For high temperatures
$k_{\rm B}T/h\gg1$, since a wide range of energies contribute in the
transport, the effect of band structure and strong energy
dependence in the density of states is washed out. Therefore the
resistivity becomes almost independent of chemical potential
$\mu$.

\begin{figure}
\includegraphics[width=8cm]{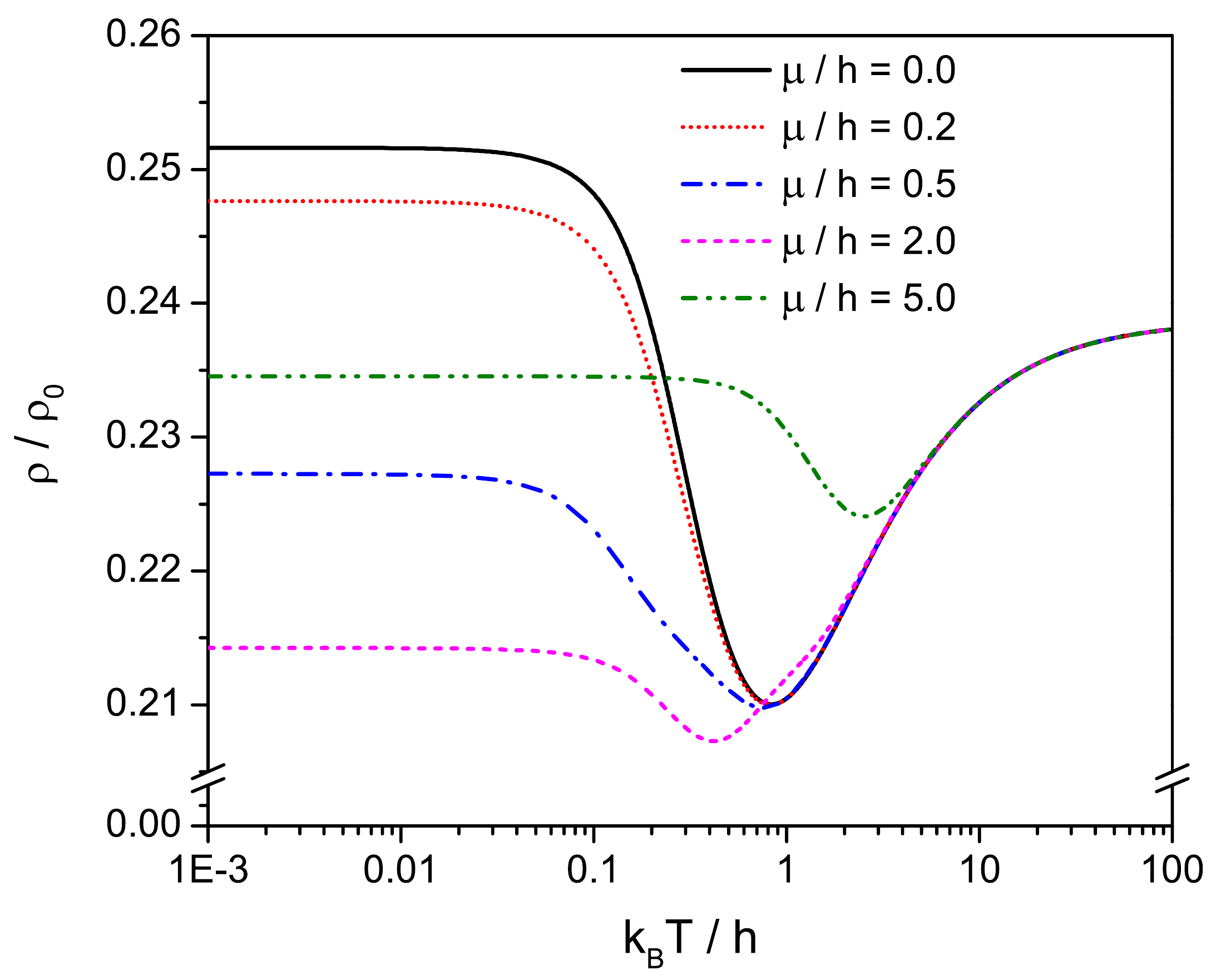}
\caption{(Color online) The resistivity versus temperature for different dopings ($\mu/h$) of magnetized graphene. The dependence exhibits a minimum
at $k_{B}T\sim h$.}\label{fig3}
\end{figure}

The main result of this paper can be seen in Fig. \ref{fig3} where the
temperature dependence of the resistivity is shown for different
chemical potentials. In general we see that the resistivity passes
through a minimum around $k_{\rm B}T/h=1$.
This behavior, at first glance, is the reminiscent of famous Kondo
effect, although the underlying physics are completely different.
In the case of the Kondo effect, strong screening of the
impurity spins at low temperatures leads to a strongly correlated many-body system,
which results in a profound scattering from magnetic impurities.
Then the resistivity starts to increase logarithmically by
decreasing the temperature below the so-called Kondo temperature
$T_{\rm K}$. Here, on the other hand, the minimum in the resistivity originates from the fact that
around $k_{\rm B}T\sim h$ the states with energies $\varepsilon\sim h$ start to contribute effectively in transport. These states, as discussed above are relatively better conducting than other states since spin-flip scatterings are suppressed for them. Therefore the conductivity increases by increasing the temperature up to $k_{\rm B}T\sim h$. Further increase in the temperature leads to a wide range of energies participating in transport and thus the contribution of states with energies $\varepsilon\sim h$ becomes negligible. So the resistivity increases again for higher temperatures $k_{\rm B}T\gg h$. We should keep in mind that here, in contrast to the Kondo effect, the minimum resistivity as a function of temperature appear as a pure single particle phenomenon and no many-body effect is considered.
Kondo physics differs from the current phenomena in several other aspects too. Namely, the Kondo temperature scales with the strength of coupling between the spin of itinerant electrons and the magnetic moment of impurities $J$, while here the minimum resistivity occurs at $T\sim h/k_{\rm B}$. Antiferromagnetic coupling between itinerant electrons and impurities (\emph{i.e.}, $J<0$) is an essential requirement for the Kondo effect, while the sign of $J$ is irrelevant in our system. Moreover, the resistivity diverges at $T\to 0$ in the Kondo effect but it saturates to a constant value here.

\section{Conclusion}\label{sec:concl}
In conclusion, we have studied the effects of magnetic impurities on electronic transport in a magnetized graphene
using the semiclassical Boltzmann theory. Taking into account both spin reversing and spin
conserving scattering, we have obtained that the temperature dependence of the resistivity
exhibits a minimum due to spin-flip induced transitions of electrons between exchange split spin sub-bands.
This effect is the direct result of the gapless Dirac spectrum of graphene in which the  density of states in the conduction and valance
bands declines linearly with varying the energy toward the Dirac point with a vanishing DOS.
The amplitude of the obtained minimum resistivity is determined by the strength of the coupling between spin of electrons and spin
of magnetic impurities, while its temperature does not depend on the coupling strength, but rather is of order of
the splitting energy.

\acknowledgments
All authors are grateful to Hosein Cheraghchi for useful discussions at the early stages of the work.
M. Z. thanks the Associate Office of ICTP in Trieste for the hospitality and support during his visit to this institute.

\end{document}